\documentclass{article}
\usepackage{amsmath,amssymb,mathrsfs,amsthm}

\theoremstyle{plain}
\newtheorem{thm}{Theorem}
\newtheorem{prop}{Proposition}[section]

\newtheorem{lemma}[prop]{Lemma}

\theoremstyle{definition}

\newtheorem{example}[prop]{Example}
\newtheorem{asm}{Assumption}

\theoremstyle{remark}
\newtheorem{remark}[prop]{Remark}

\newcommand{\RR}{\mathbb{R}}
\newcommand{\CC}{\mathbb{C}}
\newcommand{\ZZ}{\mathbb{Z}}
\newcommand{\Cinf}{C^{\infty}}
\newcommand{\ve}{\varepsilon}

\begin{document}

\title{Complex absorbing potential method for Stark resonances}
\author{Kentaro Kameoka}

\date{}

\maketitle

\begin{abstract}
We characterize the resonances of Stark Hamiltonians by the complex absorbing potential method. 
Namely, we prove that the Stark resonances are the limit points of complex eigenvalues of the
Stark Hamiltonian with a quadratic complex absorbing potential when the absorbing coefficient tends to zero.
The proof employs the complex distortion outside a cone introduced in the previous work by the author. 
Potentials with local singularities such as the Coulomb potential are allowed as perturbations. 
\end{abstract}

\section{Introduction}
In this paper, we prove that the complex absorbing potential method is valid for Stark resonances.
We study the Stark Hamiltonian
\[ P=-\Delta +x_1+V(x), \]
where $V(x)\in C^{\infty}(\mathbb{R}^n; \mathbb{R})$ is a non-globally analytic potential. 

We first recall the definitions of resonances following~\cite{K}. 
For $K>0$ and $\rho\in \RR$, 
we set the cone $C(K, \rho)=\{x \in \mathbb{R}^n | |x'| \le K(x_1+\rho)\}$, 
where $x'=(x_2, \dots, x_n)$. We denote its complement by $C(K, \rho)^c$. 
We denote the set of smooth functions which is bounded with all its derivatives by $C_b^{\infty}$. 

\begin{asm}\label{asm-potential}
The potential $V$ is decomposed as $V=V_1+V_{\mathrm{sing}}$, where $V_1$ and $V_{\mathrm{sing}}$ satisfy 
the following.

\noindent
(1). The smooth part $V_1(x)\in C_b^{\infty}(\mathbb{R}^n; \mathbb{R})$ has an analytic continuation  
to the region $\{x\in \mathbb{C}^n|\,\mathrm{Re}x\in C(K_0, \rho_0)^c, |\mathrm{Im}x|<\delta_0\}$ 
for some $\rho_0\in \mathbb{R}, K_0>0$ and $\delta_0>0$, 
and $\partial V(x)$ goes to zero when $|\mathrm{Re}x|\to \infty$ in this region.

\noindent
(2). The singular part $V_{\mathrm{sing}}\in L_{\mathrm{comp}}^2(\mathbb{R}^d; \mathbb{R})$ is 
$-\Delta$-bounded with relative bound $0$.
\end{asm}

The outgoing resolvent is denoted by $R_{+}(z)=(z-P)^{-1}$ for $\mathrm{Im}z>0$.
\begin{thm}\label{thm-0}
Suppose Assumption 1 holds.  
Then for any $\chi_1, \chi_2 \in L_{\mathrm{comp}}^{\infty}(\mathbb{R}^n)$ 
such that $\chi_j=1$ near $\mathrm{supp}\,V_{\mathrm{sing}}$, the cutoff resolvent 
$\chi_1 R_{+}(z) \chi_2 \mspace{7mu}, (\mathrm{Im}z>0)$, has a meromorphic continuation to 
$\{z|\,\mathrm{Im}z>-\delta_0\}$ with finite rank poles. 
The pole $z$ is called a resonance and the multiplicity is defined by 
\[m_z=\mathrm{rank}\frac{1}{2\pi i}\oint_z \chi_1 R_{+}(z) \chi_2 dz .\]
The set of resonances is independent of the choices of $\chi_1$,\,$\chi_2$ 
including multiplicities and denoted by $\mathrm{Res}(P)$. 
\end{thm}

We prove Theorem~1 using the complex distortion outside a cone introduced in \cite{K}. 
In \cite{K} Theorem~\ref{thm-0} was proved when $V_{\mathrm{sing}}=0$. 
The modification for the case $V_{\mathrm{sing}}\not=0$ is presented in the appendix. 
Although Theorem~\ref{thm-0} can be proved based on the complex distortion on a half space 
(\cite[Chapter~23]{HS}), the proof of Theorem~\ref{thm-1} below based on the half space distortion seems 
to be more difficult than ours based on complex distortion outside a cone. 

We next discuss the complex absorbing potential method. 

\begin{asm}\label{asm-potential-2}
In addition to Assumption~\ref{asm-potential}, the following hold.

\noindent
(1). The smooth part satisfies $\lim_{|x|\to \infty}V_1(x)=0$.  

\noindent
(2). The singular part $V_{\mathrm{sing}}$ is $-\Delta$-compact. 
\end{asm}
We note that Assumption~\ref{asm-potential-2}.(1) for $x\in \RR^n$ implies 
that for $\mathrm{Re}x\in C(K_0, \rho_0)^c$, $|\mathrm{Im}x|<\delta_0$ 
by Assumption~\ref{asm-potential}.(1).

We set 
\[
P_{\ve}=P-i\ve x^2
\]
for $\ve>0$.
Then $P_{\ve}$ for $\ve>0$ with $D(P_{\ve})=D(-\Delta)\cap D(x^2)$ 
has purely discrete spectrum (see Section~2). 
We set $m_z=0$ if $z\not\in \mathrm{Res}(P)$. 
Then our main result is the following.

\begin{thm}\label{thm-1}
Under Assumption~\ref{asm-potential-2} and the above notation,  
\[
\lim_{\ve \to 0+}\sigma_d (P_{\ve})=\mathrm{Res}(P) 
\]
in the sense that for any $z \in \{z\in \CC|\, \mathrm{Im}z>-\delta_0\}$ 
there exists $\gamma_0>0$ such that 
for any $0<\gamma<\gamma_0$ there exists $\ve_0>0$ 
such that for any $0<\ve<\ve_0$, 
\[ \# \bigl(\sigma_d(P_{\ve})\cap B(z, \gamma)\bigr)=m_z,\]
where $B(z, \gamma)=\{w \in \mathbb{C}|\, |w-z|\le \gamma\}$.
\end{thm}

\begin{example}
Consider the Coulomb potential $V(x)=\sum_{i=1}^{N}\frac{e_i}{|x-R_i|}$ on $\mathbb{R}^n$, $(n\ge 3)$. 
Since $(|\mathrm{Re}x|^2-|\mathrm{Im}x|^2+2i \mathrm{Re}x \cdot \mathrm{Im}x)^{-1/2}$ is well-defined for
$|\mathrm{Re}x|>|\mathrm{Im}x|$, this potential satisfies the Assumption~\ref{asm-potential-2} 
with arbitrary large $\delta_0$ if we take $\rho_0$ and $K$ large. 
Thus the resonances for $-\Delta+ x_1+ 
\sum_{i=1}^{N}\frac{e_i}{|x-R_i|}$ are well-defined on the whole complex plane 
and they are characterized as limit points of discrete eigenvalues of 
$-\Delta+ x_1+ 
\sum_{i=1}^{N}\frac{e_i}{|x-R_i|}-i\ve x^2$ when $\ve\to 0+$.
\end{example}

Resonances correspond to quasi-steady states and are closely related to scattering theory. 
The resonance theory is a rich branch of spectral theory (see~\cite{DZ2}). 
The resonances for Stark Hamiltonians were studied by many authors  
(for instance, \cite{H}, \cite{H2}, \cite{HeSi}, \cite{Si}, \cite{Si2}, \cite{W2}, \cite{W3}). 
Complex absorbing potential method was introduced in physical chemistry (\cite{RM}, \cite{SeMi}). 
The mathematical justification was given by Zworski~\cite{Z2} for compactly supported potentials. 
This was extended to several settings (see~\cite{KN}, \cite{X1}, \cite{X2}, \cite{X3}). 
An analogous result holds for Policott-Ruelle resonances (see~\cite{DZ1}).

This paper is organized as follows. 
In Section~2, we recall the complex distortion outside a cone and discuss its 
generalization to the Stark Hamiltonian with complex absorbing potential. 
In Section~3, we prove Theorem~\ref{thm-1} 
by constructing an approximate resolvent of the free Stark distorted operator. 
In Section~4, we prove technical lemmas used in the proof of Theorem~\ref{thm-1}. 
In the appendix, we discuss the modifications needed for including local singularities of the potential 
and prove Theorem~\ref{thm-0}.

\section{Complex distorted operator and complex absorbing potential}
We first recall the complex distortion outside a cone introduced in \cite{K}. 
We take any $K>K_0$ and sufficiently large $\rho>0$ and deform $P$ in $C(K, \rho)^c$. 
Take a convex set $\widetilde{C}(K, \rho)$ which has a smooth boundary 
such that $\widetilde{C}(K, \rho)$ is rotationally symmetric with respect to $x'$ and 
$\widetilde{C}(K, \rho)=C(K,\rho)$ in $x_1>-\rho+1$.
We define $F=-(1+K^{-2})^{\frac{1}{2}}\mathrm{dist}\left(\bullet, \widetilde{C}(K, \rho)\right)\ast \phi$, 
where $\phi \in C_c^{\infty}(\mathbb{R}^n)$, 
$\mathrm{supp}\,\phi \subset \{|x|<1\}$,  $\phi \ge 0$ and $\int\phi(x)dx=1$. 
We also set $v(x)=(v_1(x), \dots, v_n(x))=\partial F(x)\in C_b^{\infty}(\mathbb{R}^n; \mathbb{R}^n)$. 
We next set $\Phi_{\theta}(x)=x+\theta v(x)$. 
This is a diffeomorphism for real $\theta$ with small $|\theta|$. 
We also write $x_{\theta}=\Phi_{\theta}(x)$.
We set $U_{\theta}f(x)=\left(\mathrm{det}\Phi_{\theta}'(x)\right)^{\frac{1}{2}}f(\Phi_{\theta}(x))$, 
which is unitary on $L^2(\mathbb{R}^n)$.
We define the distorted operator $P_{\theta}=U_{\theta}P U_{\theta}^{-1}$. 
Then 
\begin{align*}
P_{\theta}
&=-\sum_{i,j} \partial_i g_{\theta}^{ij}\partial_j
+r_{\theta}(x)+x_1+\theta v_1+V(x_{\theta}),
\end{align*}
where $(g_{\theta}^{ij})=(\Phi_{\theta}')^{-2}$, $g_{\theta}=\mathrm{det}(\Phi_{\theta}')^{2}$ 
and $r_{\theta}=-\sum_{i,j}g_{\theta}^{-\frac{1}{4}}
(\partial_i(g_{\theta}^{\frac{1}{2}}g_{\theta}^{ij}\partial_j g_{\theta}^{-\frac{1}{4}}))$.
For $\mathrm{Im}\theta<0$, we have $\mathrm{Im}(-\sum_{i,j} \partial_i g_{\theta}^{ij}\partial_j)\le 0$ 
in the form sense (\cite[Lemma~2.1]{K}). We have $v_1(x)\ge 1$ on $C(K, \rho+1)^c$. 
The function $r_{\theta}(x)\in \Cinf_b(\RR^n)$ satisfies 
$|r_{\theta}(x)|\le C\langle x \rangle^{-1}$ if $\mathrm{dist}(x, \partial\widetilde{C}(K, \rho))>1$ 
since we have $|\partial^{\alpha}v_j(x)|\le C_{\alpha}\langle x \rangle^{-1}$ 
if $\mathrm{dist}(x, \partial\widetilde{C}(K, \rho))>1$ and $|\alpha|\ge 1$.

The $P_{\theta}$ is an analytic family of closed operators for $\theta$ with 
$|\mathrm{Im}\theta|<\delta_0(1+K^{-2})^{-\frac{1}{2}}$ and $|\mathrm{Re}\theta|$ small. 
We also have $P_{\theta}^*=P_{\bar{\theta}}$. 
Moreover $P_{\theta}$ with $\mathrm{Im}\theta <0$ has discrete spectrum in 
$\{\mathrm{Im}z>\mathrm{Im}\theta \}$. 
Resonances for $P$ coincide with discrete eigenvalues of $P_{\theta}$ in this region.  
These facts are proved in \cite[Section~2]{K} for $V_{\mathrm{sing}}=0$ 
and in the appendix for $V_{\mathrm{sing}}\not=0$. 
We set 
\[
S(m)=\{a(x, \xi)\in \Cinf(\RR^{2n})|\, |\partial_{x, \xi}^{\alpha}a(x, \xi)|\le C_{\alpha}m(x, \xi)\}
\]
with the corresponding class of pseudodifferential operators denoted by $\mathrm{Op}S(m)$. 
We recall that $P_{\ve}=P-i\ve x^2$. 
We also deform $P_{\ve}$ to obtain $P_{\ve, \theta}=U_{\theta}P_{\ve}U_{\theta}^{-1}$. 
By the ellipticity in the symbol class $S(1+x^2+\xi^2)$, 
it is easier to see analogous properties for $P_{\ve, \theta}$ with $\ve>0$. 
Namely, $P_{\ve, \theta}$ with $D(P_{\ve, \theta})=D(-\Delta)\cap D(x^2)$ 
is an analytic family of closed operators for $\theta$ with 
$|\mathrm{Im}\theta|<\delta_0(1+K^{-2})^{-\frac{1}{2}}$ and $|\mathrm{Re}\theta|$ small. 
We also have $P_{\ve, \theta}^*=P_{-\ve, \bar{\theta}}$.
The deformed operator $P_{\ve, \theta}$ with $\ve>0$ has purely discrete spectrum on the whole complex plane 
and the eigenvalues are independent of $\theta$ including multiplicities. 

The following lemma simplifies the proof of Theorem~\ref{thm-1}. 
\begin{lemma}\label{vector-field}
\[
x\cdot v(x)\le 0 \hspace{0.2cm}\text{for any}\hspace{0.2cm} x\in \RR^n. 
\]
\end{lemma}
\begin{proof}
By the rotational symmetry with respect to $x'$, we may assume that $n=2$ and $x'=x_2\ge 0$. 
We first assume that $x_1\le 0$. 
Then we have $v_2(x)\le 0$ by the construction of $v$. 
Since $v_1(x)\ge 0$ everywhere, it follows that $x\cdot v(x)\le 0$. 
We next assume that $x_1>0$. 
Then we have $v_2(x)=-K^{-1}v_1(x)$ by the construction of $v$. 
If $x_2>Kx_1$, we then have 
\[
x\cdot v(x)=x_1 v_1(x)+x_2 v_2(x)=v_1(x)(x_1-K^{-1}x_2)\le 0
\]
since $v_1(x)\ge 0$. 
If $x_2\le Kx_1$, then $v(x)=0$ by the construction of $v$, which completes the proof.
\end{proof}
A difficulty of the analysis of Stark Hamiltonians is the lack of the global ellipticity of its full symbol 
due to $\lim_{x_1\to -\infty}(V(x)+x_1)=-\infty$. 
That is, while the natural symbol class for $|\xi|^2+x_1+V(x)$ is $S(|\xi|^2+|x_1|+1)$, 
there is no $z$ such that $||\xi|^2+x_1+V(x)-z|\gtrsim (|\xi|^2+|x_1|+1)$. 
Lemma~\ref{vector-field} implies that 
\[
\mathrm{Re}(-i\ve x_{-i\delta}^2)=-2\delta\ve x\cdot v(x)\ge 0
\]
for $\ve, \delta>0$. 
Thus the distorted complex absorbing potential does not cause additional difficulty concerning 
the lack of the global ellipticity of $\mathrm{Re}P_{\ve, \theta}$.

\section{Justification of complex absorbing potential method for Stark Hamiltonians}

\subsection{Resolvent estimate for the free distorted operator with complex absorbing potential}
For simplicity, we write $P_0=P$ and $P_{0, \theta}=P_{\theta}$.
For a given $\Omega \Subset \{z|\,\mathrm{Im}z>-\delta_0\}$, 
we fix $0<\delta<\delta_0$ such that 
$\Omega \Subset \{z|\,\mathrm{Im}z>-\delta\}$ 
and set $\theta=-i\delta$. 
Then $P_{\ve, \theta}$ is well-defined for this $\theta$ 
if we take $K>0$ large enough in the definition of the complex distortion. 
Stark resonances coincide with eigenvalues of $P_{0, \theta}$ in $\Omega$.
We denote $P_{\ve, \theta}$ with $V\equiv 0$ by $Q_{\ve, \theta}$. 
We have the following resolvent estimate for 
$Q_{\ve, \theta}$, which is uniform with respect to small $\ve>0$.
\begin{prop}\label{main-estimate}
There exists $C>0$ such that  
\[
\|(Q_{\ve, \theta}-z)^{-1}\|_{L^2(\RR^n) \to L^2(\RR^n)}\le C
\]
for small $\ve>0$ and $z\in \Omega$. 
\end{prop}
\begin{remark}
We have 
\begin{align*}
Q_{\ve}&=-\Delta+x_1-i\ve x^2=\Delta-i\ve (x_1+\frac{i}{2\ve})^2-i\ve x'^2-\frac{i}{4\ve}.
\end{align*}
The eigenfunctions for this operator are given by a suitable complex coordinate transform of 
those for the harmonic oscillator. Then we have 
\[ 
\sigma(Q_{\ve})=\bigl\{\ve^{1/2}e^{-\pi i/4}(2|\alpha|+n)-\frac{i}{4\ve}\bigm|\, \alpha\in \ZZ^n_{\ge 0}\bigr\} 
\]
including multiplicities. 
This diverges to infinity when $\ve \to 0+$. 
Since $\sigma(Q_{\ve})=\sigma(Q_{\ve, \theta})$, we conclude that $(Q_{\ve, \theta}-z)^{-1}$ 
exists for small $\ve>0$ and $z\in \Omega$. 
Proposition~\ref{main-estimate} claims the uniform resolvent estimate. 
In \cite{Z2}, the complex absorbing potential method was justified in the region $\{z|\, -\pi/4<\arg z\le 0\}$ 
since $\sigma(-\Delta-i\ve x^2)=\{\ve^{1/2}e^{-\pi i/4}(2|\alpha|+n)|\, \alpha\in \ZZ^n_{\ge 0}\}$. 
In the Stark Hamiltonian case, there is no such restriction.   
\end{remark}
Proposition~~\ref{main-estimate} and its proof is crucial in the proof of Theorem~\ref{thm-1}.
To prove this proposition, we construct an approximation of $(Q_{\ve, \theta}-z)^{-1}$ as in 
\cite[Section~3]{Z2}. 
We first note that $(Q_{0, \theta}-z)^{-1}$ exists for $z\in \Omega$ by the non-existence of 
the free Stark resonances (see~\cite[Corollary~2.2]{K}). 
Thus $(Q_{0, \theta}-z)^{-1}$ will approximate $(Q_{\ve, \theta}-z)^{-1}$ on a compact set if 
$\ve$ is sufficiently small. We next construct an approximation of $(Q_{\ve, \theta}-z)^{-1}$ 
near infinity. 
Take $\chi\in \Cinf_{c}(\RR^n; [0, 1])$ such that $\chi=1$ near $x=0$. 
We set 
\[
Q_{\ve, \theta}^R=Q_{\ve, \theta}-iR\chi(x/R), \quad R\gg 1.
\]
\begin{lemma}\label{estimate-1}
There exist $R_0>1$, $\tilde{\ve}>0$ and $C>0$ such that 
\[
\|(Q_{\ve, \theta}^R-z)^{-1}\|_{L^2(\RR^n) \to L^2(\RR^n)}\le C
\]
for $R>R_0$, $0<\ve\le \tilde{\ve}$ and $z\in \Omega$. 
\end{lemma}
The proof of Lemma~\ref{estimate-1} is given in subsection~\ref{proof-estimate-1}. 
Although Lemma~\ref{estimate-1} is also true for $\ve=0$ by the same proof, 
we do not need this since $(Q_{0, \theta}-z)^{-1}$ exists. 
We fix $R>R_0$. 
We set $\chi^M(x)=\chi(x/M)$, where $\chi$ is as above.  
 
Then our approximate resolvent of $Q_{\ve, \theta}$ is defined by 
\[
T_{\ve, z}^M=\chi^M(Q_{0, \theta}-z)^{-1}+(1-\chi^M)(Q_{\ve, \theta}^R-z)^{-1}.
\]
Lemma~\ref{estimate-1} implies that $T_{\ve, z}^M$ is uniformly bounded for 
$0< \ve \le \tilde{\ve}$, $M>1$ and $z\in \Omega$. 
We set $(Q_{\ve, \theta}-z)T_{\ve, z}^M=1+E_{\ve, z}^M$. 
Then a simple calculation shows that 
\[
E_{\ve, z}^M=[Q_{\ve, \theta}, \chi^M](Q_{0, \theta}-z)^{-1}
-[Q_{\ve, \theta}, \chi^M](Q_{\ve, \theta}^R-z)^{-1}
-\chi^Mi\ve x_{\theta}^2(Q_{0, \theta}-z)^{-1} 
\]
for $M\gg R$. 
To estimate this, we give the following lemma. 
We set $\widetilde{Q}_{\ve}=Q_{\ve, \theta}^R$ for $0<\ve\le \tilde{\ve}$ and 
$\widetilde{Q}_{0}=Q_{0, \theta}$. 
For any $\tilde{\chi}\in \Cinf_b(\RR)$, we set $\tilde{\chi}^M(x)=\tilde{\chi}(x_1/M)$. 

\begin{lemma}\label{estimate-2}
Take any $\tilde{\chi}\in \Cinf_b(\RR)$ such that $\tilde{\chi}(x_1)=0$ for $-x_1\gg 1$. 
Then there exists $C>0$ such that 
\[
\|\tilde{\chi}^M(\widetilde{Q}_{\ve}-z)^{-1}\|_{L^2(\RR^n)\to H^k(\RR^n)}\le CM^{k/2}  
\]
for $0\le k \le 2$, $M>1$, $z\in \Omega$ and $0\le \ve \le \tilde{\ve}$. 
\end{lemma}

The proof of Lemma~\ref{estimate-2} is given in subsection~\ref{proof-estimate-2}. 
Since the commutator improves $M^{-1}$, Lemma~\ref{estimate-2} with $k=1$ implies that 
\[
\|E_{\ve, z}^M\|_{L^2(\RR^n)\to L^2(\RR^n)}=\mathcal{O}(M^{-1/2})+\mathcal{O}_M(\ve).
\]
If we take $M$ large and then take $\ve>0$ small, we have 
$\|E_{\ve, z}^M\|_{L^2(\RR^n)\to L^2(\RR^n)}<1/2$ for $z \in \Omega$. 
Thus the Neumann series argument implies that $T_{\ve, z}^M(1+E_{\ve, z}^M)^{-1}$ is a right inverse of 
$Q_{\ve, \theta}-z$. Since the adjoint $Q_{-\ve, \bar{\theta}}-\bar{z}$ also has a right inverse by 
the same argument, we conclude that $(Q_{\ve, \theta}-z)^{-1}$ exists and 
is equal to $T_{\ve, z}^M(1+E_{\ve, z}^M)^{-1}$ on $L^2(\RR^n)$. 
Then the Proposition~\ref{main-estimate} follows from the uniform boundedness of $T_{\ve, z}^M$.

\subsection{Proof of convergence to Stark resonances}
\begin{proof}[Proof of Theorem~\ref{thm-1}]
The strategy of the proof is the same as \cite[Section~5]{Z2} (see also \cite[subsection~2.3]{KN}). 
By Proposition~\ref{main-estimate}, we write 
\[
P_{\ve, \theta}-z=(1+V_{\theta}(Q_{\ve, \theta}-z)^{-1})(Q_{\ve, \theta}-z),
\]
where $V_{\theta}(x)=V(x_{\theta})$. 
Note that $\lim_{x\to \infty} V_{\theta}(x)=0$ and $V_{\theta}$ is $-\Delta$-compact by our assumption. 
By approximating $V_{\theta}$ by compactly supported $-\Delta$-compact functions, 
Lemma~\ref{estimate-2} with $k=2$ implies that $V_{\theta}(Q_{\ve, \theta}-z)^{-1}$ is a compact operator and 
thus $1+V_{\theta}(Q_{\ve, \theta}-z)^{-1}$ is a Fredholm operator. 
Thus the same arguments as in \cite{Z2}, \cite[subsection~2.3]{KN}  
based on the analytic Fredholm theory and the Gohberg-Sigal theory 
implies that Theorem~\ref{thm-1} follows if we prove 
\begin{equation}\label{limit-eq}
\lim_{\ve \to 0+}\|V_{\theta}(Q_{\ve, \theta}-z)^{-1}
-V_{\theta}(Q_{0, \theta}-z)^{-1}\|_{L^2(\RR^n)\to L^2(\RR^n)}=0
\end{equation}
uniformly for $z\in \Omega$. 
While \cite{Z2}, \cite{KN} employ the resolvent equation, we instead use the construction of approximate 
resolvent in subsection~3.1 since the estimate of the distorted Stark resolvent with the weight $x_{\theta}^2$ 
is not easy. 
By approximating $V_{\theta}$ by compactly supported functions and using Proposition~\ref{main-estimate}, 
it is enough to prove (\ref{limit-eq}) with 
$V_{\theta}$ replaced by some $\widetilde{V_{\theta}}$, where 
$\widetilde{V_{\theta}}\in L^2_{\mathrm{comp}}(\RR^n)$ and $\widetilde{V_{\theta}}$ is $-\Delta$-compact. 
By subsection~3.1, we have 
\begin{align*}
\widetilde{V_{\theta}}(Q_{\ve, \theta}-z)^{-1}-\widetilde{V_{\theta}}(Q_{0, \theta}-z)^{-1}&
=\widetilde{V_{\theta}}(T_{\ve, z}^M(1+E_{\ve, z}^M)^{-1}-(Q_{0, \theta}-z)^{-1})\\
&=\widetilde{V_{\theta}}(Q_{0, \theta}-z)^{-1}((1+E_{\ve, z}^M)^{-1}-1).
\end{align*}
Here we took $M>1$ large enough in the definition of $T_{\ve, z}^M$ and used fact that 
$\widetilde{V_{\theta}}T_{\ve, z}^M=\widetilde{V_{\theta}}(Q_{0, \theta}-z)^{-1}$ for $0<\ve\le \tilde{\ve} $ 
since $\widetilde{V_{\theta}}(1-\chi^M)=0$. 
Note that $\widetilde{V_{\theta}}(Q_{0, \theta}-z)^{-1}$ is independent of $\ve > 0$ and $M>1$ 
and is a bounded operator on $L^2(\RR^n)$ by 
Lemma~\ref{estimate-2} with $k=2$, the $-\Delta$-boundedness of $\widetilde{V_{\theta}}$ 
and the compactness of $\mathrm{supp}\, \widetilde{V_{\theta}}$. 
Then we have 
\begin{align*}
&\|\widetilde{V_{\theta}}(Q_{\ve, \theta}-z)^{-1}-\widetilde{V_{\theta}}(Q_{0, \theta}-z)^{-1}\|
_{L^2(\RR^n)\to L^2(\RR^n)}\\
&\lesssim \|(1+E_{\ve, z}^M)^{-1}-1\|_{L^2(\RR^n)\to L^2(\RR^n)}.
\end{align*}
Recall that 
\[
\|E_{\ve, z}^M\|_{L^2(\RR^n)\to L^2(\RR^n)}=\mathcal{O}(M^{-1/2})+\mathcal{O}_M(\ve).
\]
Thus if we take large $M>1$ and then take small $\ve>0$, 
the operator $(1+E_{\ve, z}^M)^{-1}$ is close to the identity operator in the 
operator norm. 
This completes the proof of Theorem~\ref{thm-1}.
\end{proof}

\section{Proofs of technical lemmas}
In this section, we prove two lemmas in Section~3. 
The notation is the same as in Section~3. 
Take $w \in  C^{\infty}(\mathbb{R}^n ; \mathbb{R}_{\ge 1})$ depending only on $x_1$ and 
$w=|x_1|$ for $x_1 \le -2$ and $w=1$ for $x_1 \ge -1$. 
\subsection{Proof of Lemma~\ref{estimate-1}}\label{proof-estimate-1}
We take small $c_0>0$ and $\tilde{\chi}_1, \tilde{\chi}_2, \tilde{\chi}_3, \tilde{\chi}_4 
\in \Cinf(\RR; [0,1])$ such that 
$\tilde{\chi}_1(x_1)=1$ for $x_1> 5c_0$, $\tilde{\chi}_1(x_1)=0$ for $x_1<4c_0$, 
$\tilde{\chi}_2(x_1)=1$ for $x_1<5c_0 $, $\tilde{\chi}_2(x_1)=0$ for $x_1>6c_0$, 
$\tilde{\chi}_3(x_1)=1$ for $|x_1-5c_0|<c_0$, $\tilde{\chi}_3(x_1)=0$ for $|x_1-5c_0|>2c_0$,  
$\tilde{\chi}_4(x_1)=1$ for $x_1<c_0$ and $\tilde{\chi}_4(x_1)=0$ for $x_1>2c_0$. 
We set $\tilde{\chi}_j^R(x)=\tilde{\chi}_j(x_1/R)$ for $j=1, 2, 3, 4$. 

We take any $u\in \Cinf_c(\RR^n)$. 
We have $\|u\|\le \|\tilde{\chi}_1^R u\|+\|\tilde{\chi}_2^R u\|$. 
We first note that 
\[
\mathrm{Re}(\tilde{\chi}_1^Ru, (Q_{\ve, \theta}^R-z)\tilde{\chi}_1^Ru)
\gtrsim R\|\tilde{\chi}_1^R u\|^2
\]
by the Stark potential. 
We also have 
\[
\mathrm{Im}(\tilde{\chi}_2^Ru, (Q_{\ve, \theta}^R-z)\tilde{\chi}_2^Ru)
\lesssim -\|\tilde{\chi}_2^R u\|^2
\]
by the complex distortion outside a cone 
and the $-iR\chi(x/R)$ term in $Q_{\ve, \theta}^R$ (if $c_0$ is sufficiently small). 
Here we assumed $0<\ve \le \tilde{\ve}$, where $\tilde{\ve}>0$ is small, to estimate the term 
$i\ve \delta^2 v(x)^2$ in $-i\ve x_{-i\delta}^2$. 
Thus we have 
\begin{align*}
\|u\|&\lesssim \sum_{j=1}^2\|(Q_{\ve, \theta}^R-z)\tilde{\chi}_j^Ru\|\\
&\lesssim \|(Q_{\ve, \theta}^R-z)u\|+
\sum_{j=1}^2\|[Q_{\ve, \theta}^R, \tilde{\chi}_j^R]u\|\\
&\lesssim \|(Q_{\ve, \theta}^R-z)u\|+
R^{-1}\|\tilde{\chi}_3^Ru\|_{H^1}.
\end{align*} 

We set  
\[
\widehat{Q}=\widehat{Q}_{\ve, z}^R
=Q_{\ve, \theta}^R-z+\tilde{\chi}_4^Rw+R\tilde{\chi}_4^R.  
\]
The estimates below are uniform with respect to $0<\ve\le \tilde{\ve}$ and $z\in \Omega$. 
We define the symbol $\widehat{q}(x, \xi; R)$ by $\widehat{Q}_{\ve, z}^R=\widehat{q}(x, D; R)$. 
We now prove that $\widehat{Q}^{-1}\in \mathrm{Op}S(\langle \xi \rangle^{-2})$ uniformly for $R\gg 1$. 
By Lemma~\ref{vector-field} and $x_1+\tilde{\chi}_4^Rw+R\tilde{\chi}_4^R \gtrsim R$, we have 
\begin{align*}
&\left|\frac{1+\xi^2+\ve  x^2 }
{\widehat{q}(x, \xi; R)}\right|,\,
\left|\frac{\partial_x\widehat{q}(x, \xi; R)}
{\widehat{q}(x, \xi; R)}\right|
\lesssim \frac{1+\xi^2+\ve  x^2 }
{|\xi^2+R-i\ve x^2|}\lesssim 1 
\end{align*}
and 
\begin{align*}
&\left|\frac{\partial_{\xi}\widehat{q}(x, \xi; R)}
{\widehat{q}(x, \xi; R)}\right|
\lesssim \frac{|\xi|}
{|\xi^2+R-i\ve x^2|}.
\end{align*}
We have 
\[
\frac{\langle\xi \rangle}{|\xi^2+R-i\ve x^2|}\lesssim R^{-1/2}
\]
by estimating it separately for $|\xi|/R^{1/2}\gg 1$ and $|\xi|/R^{1/2}\lesssim 1$. 
These imply that $\widehat{q}^{-1}(x, \xi; R)=\mathcal{O}(1)$ in 
$S((1+\xi^2+\ve x^2)^{-1})\subset S(\langle \xi \rangle^{-2})$ for $R>1$, 
$\partial_x\widehat{q}^{-1}(x, \xi; R)=\mathcal{O}(R^{-1/2})$ in $S(\langle \xi \rangle^{-1})$ and 
$\partial_{\xi}\widehat{q}^{-1}(x, \xi; R)=\mathcal{O}(R^{-1/2})$ in $S((1+\xi^2+\ve x^2)^{-1})$. 
Then a standard argument of pseudodifferential operators (see~\cite[Chapter~5]{Z})
implies that the seminorms in $S(1)$ of the symbols of 
$\widehat{q}(x, D; R)\widehat{q}^{-1}(x, D; R)-1$ and 
$\widehat{q}^{-1}(x, D; R)\widehat{q}(x, D; R)-1$ are $\mathcal{O}(R^{-1/2})$ 
since $\partial_x \widehat{q}(x, \xi; R)\in S(1+\xi^2+\ve x^2)$ and 
$\partial_{\xi} \widehat{q}(x, \xi; R)\in S(\langle \xi \rangle)$. 
We note that this estimate based on the pseudodifferential calculus in the symbol classes 
$S((1+\xi^2+\ve x^2)^{\pm 1})$ is uniform with respect to $0<\ve \le 1$ 
since we have $|(1+\xi^2+\ve x^2)|/|(1+\eta^2+\ve y^2)|\le C(1+|x-y|+|\xi-\eta|)^N$ uniformly for 
$0< \ve \le 1$ (see~\cite[Chapter~5]{Z}). 
(In these arguments, we may replace $1+\xi^2+\ve x^2$ by $1+\xi^2+\ve |x|$ if we use 
$|\partial^{\alpha}(\ve x_{\theta}^2)|\le C_{\alpha}\ve\langle x \rangle$ 
although we only need $|\partial^{\alpha}(\ve x_{\theta}^2)|\le C_{\alpha}\ve\langle x \rangle^2$.)
Then the Neumann series argument and the Beals's theorem imply that 
$\widehat{Q}^{-1}\in \mathrm{Op}S(\langle \xi \rangle^{-2})$ uniformly for $R\gg 1$. 
In particular, $\widehat{Q}^{-1}: H^{-1}(\RR^n)\to H^1(\RR^n)$ is uniformly bounded for $R\gg 1$. 

Then we have 
\[
\|\tilde{\chi}_3^Ru\|_{H^1}\lesssim \|\widehat{Q} \tilde{\chi}_3^Ru\|_{H^{-1}}.
\]
Since $\mathrm{supp}\,\tilde{\chi}_3^R\cap \mathrm{supp}\,\tilde{\chi}_4^R=\emptyset$, 
this is equal to $\|( Q_{\ve, \theta}^R-z)\tilde{\chi}_3^Ru\|_{H^{-1}}$. 
This is bounded by 
\begin{align*}
&\|( Q_{\ve, \theta}^R-z)u\|+\|[Q_{\ve, \theta}^R, \tilde{\chi}_3^R]u\|_{H^{-1}}
\lesssim\|( Q_{\ve, \theta}^R-z)u\|+R^{-1}\|u\|. 
\end{align*}
Thus we have $\|u\|\lesssim \|( Q_{\ve, \theta}^R-z)u\|$ for $R\gg 1$. 
Since $\Cinf_c(\RR^n)$ is a core for $Q_{\ve, \theta}^R$, this is true for any $u$ 
in the domain of $Q_{\ve, \theta}^R$. 
Since the adjoint $(Q_{\ve, \theta}^R-z)^*=Q_{-\ve, \bar{\theta}}-\bar{z}+iR\chi(x/R)$ 
has the same estimate, we conclude that 
$( Q_{\ve, \theta}^R-z)^{-1}$ exists on $L^2(\RR^n)$ and 
$\|( Q_{\ve, \theta}^R-z)^{-1}\|_{L^2\to L^2}\le C$, which completes the proof of 
Lemma~\ref{estimate-1}.

\subsection{Proof of Lemma~\ref{estimate-2}}\label{proof-estimate-2}
We note that Lemma~\ref{estimate-2} for $M>1$ follows from that for $M\gg 1$. 
The existence of $(Q_{0, \theta}-z)^{-1}$ and Lemma~\ref{estimate-1} imply that 
Lemma~\ref{estimate-2} is valid for $k=0$ (without $\tilde{\chi}^M$). 
Thus it is enough to prove the case of $k=2$ by the interpolation argument. 

Take $\tilde{\chi}_5. \tilde{\chi}_6\in \Cinf_b(\RR; [0, 1])$ 
such that $\tilde{\chi}_5=1$ near $\mathrm{supp}\, \tilde{\chi}$ 
and $\tilde{\chi}_6=1$ near $\mathrm{supp}\, \tilde{\chi}_5$. 
Set $\tilde{\chi}_j^M(x)=\tilde{\chi}_j (x_1/M)$. 
We fix large $C_1>0$ and set 
\[
A=A_{\ve, z}^M=\widetilde{Q}_{\ve}+(1-\tilde{\chi}_6^M) w+C_ 1M-z.
\]
The estimates below are uniform with respect to $0\le\ve\le \tilde{\ve}$ and $z\in \Omega$. 
We define the symbol $a(x, \xi; M)$ by $A_{\ve, z}^M=a(x, D; M)$. 
We now prove that $A^{-1}\in \mathrm{Op}S(\langle \xi \rangle^{-2})$ uniformly for $M\gg 1$. 
By Lemma~\ref{vector-field} and 
$\xi^2+x_1+(1-\tilde{\chi}_6^M) w+C_1M\gtrsim \xi^2+M$ for $C_1\gg 1$, we have  
\begin{align*}
&\left|\frac{1+\xi^2+\ve  x^2 }
{a(x, \xi; R)}\right|,\,\left|\frac{\partial_x a(x, \xi; M)}
{a(x, \xi; M)}\right|
\lesssim \frac{1+\xi^2+\ve  x^2 }
{|\xi^2+M-i\ve x^2|}\lesssim 1 
\end{align*}
and 
\begin{align*}
&\left|\frac{\partial_{\xi}a(x, \xi; M)}
{a(x, \xi; M)}\right|
\lesssim \frac{|\xi|}
{|\xi^2+M-i\ve x^2|}.
\end{align*}

Then the same argument as in the proof of Lemma~\ref{estimate-1} with $R$ replaced by $M$ and 
$q$ replaced by $a$ implies that 
$A^{-1}\in \mathrm{Op}S(\langle \xi \rangle^{-2})$ uniformly for $M\gg 1$. 
In particular, $A^{-1}: H^k(\RR^n)\to H^{k+2}(\RR^n)$ is uniformly bounded for any $k\in \RR$ 
for $M\gg 1$. 

To estimate $\|\tilde{\chi}^M(\widetilde{Q}_{\ve}-z)^{-1}\|_{L^2\to H^2}$, we decompose 
\[
\tilde{\chi}^M(\widetilde{Q}_{\ve}-z)^{-1}=\tilde{\chi}^MA^{-1}\tilde{\chi}_5^M A (\widetilde{Q}_{\ve}-z)^{-1}+ 
\tilde{\chi}^MA^{-1}(1-\tilde{\chi}_5^M) A (\widetilde{Q}_{\ve}-z)^{-1}.
\]
Since $\tilde{\chi}_5^M(1-\tilde{\chi}_6^M)=0$, we estimate the first term as  
\begin{align*}
&\|\tilde{\chi}^MA^{-1}\tilde{\chi}_5^M A (\widetilde{Q}_{\ve}-z)^{-1}\|_{L^2\to H^2}\\
&\lesssim \|A^{-1}\|_{L^2\to H^2}\cdot \|(\widetilde{Q}_{\ve}+C_1M-z)
(\widetilde{Q}_{\ve}-z)^{-1}\|_{L^2\to L^2}\\
&\lesssim M, 
\end{align*}
where we also used the fact that 
$(\widetilde{Q}_{\ve}-z)^{-1}:L^2(\RR^n)\to L^2(\RR^n)$ is uniformly bounded (Lemma~\ref{estimate-1}). 
Since $\tilde{\chi}^M(1-\tilde{\chi}_5^M)=0$, we estimate the second term as 
\begin{align*}
&\|\tilde{\chi}^MA^{-1}(1-\tilde{\chi}_5^M) A (\widetilde{Q}_{\ve}-z)^{-1}\|_{L^2\to H^2}\\
&=\|A^{-1}[\tilde{\chi}^M, A]A^{-1}(1-\tilde{\chi}_5^M) A (\widetilde{Q}_{\ve}-z)^{-1}\|_{L^2\to H^2}\\
&=\|A^{-1}[\tilde{\chi}^M, A]A^{-1}[\tilde{\chi}_5^M, A] (\widetilde{Q}_{\ve}-z)^{-1}\|_{L^2\to H^2}\\
&\lesssim\|A^{-1}\|_{L^2\to H^2}\cdot \|[\tilde{\chi}^M, A]\|_{H^1\to L^2}\cdot 
\|A^{-1}\|_{H^{-1}\to H^1}\cdot \|[\tilde{\chi}_5^M, A]\|_{L^2\to H^{-1}}\\
&\lesssim M^{-2}.
\end{align*}
Here we used facts that $[\tilde{\chi}^M, A]A^{-1}A(1-\tilde{\chi}_5^M)=[\tilde{\chi}^M, A](1-\tilde{\chi}_5^M)=0$ and 
that $[\tilde{\chi}^M, A]$ and $[\tilde{\chi}_5^M, A]$ 
are first order differential operators with $\mathcal{O}(M^{-1})$ coefficients. 
Summing up these, we have $\|\tilde{\chi}^M(\widetilde{Q}_{\ve}-z)^{-1}\|_{L^2\to H^2}\lesssim M$ for 
$M\gg 1$, which completes the proof of Lemma~\ref{estimate-2}.

\begin{remark}
In fact, we can prove Theorem~\ref{thm-1} without relying on the specific property of the vector field $v(x)$ 
in Lemma~\ref{vector-field}. 
If we do not use Lemma~\ref{vector-field}, we replace $w(x_1)$ with $\langle x \rangle$ in the proofs of 
Lemma~\ref{estimate-1} and Lemma~\ref{estimate-2}. 
We also replace $\tilde{\chi}_j$ with cutoffs near 
$C(K_j, \rho_j)$, $C(K_j, \rho_j)^c$ or $\partial C(K_j, \rho_j)$ for suitable $K_j>0$ and $\rho_j\in \RR$.  
Then Lemma~\ref{estimate-1} and Lemma~\ref{estimate-2} are proved if 
$\tilde{\chi}$ in Lemma~\ref{estimate-2} is replaced by a cutoff near $C(\tilde{K}, \tilde{\rho})$ 
for any $\tilde{K}>0$ and $\tilde{\rho}>0$. 
\end{remark}

\appendix 

\section{Local singularities}
In this appendix, 
we present modifications to include local singularities of the potential. 
In particular, we prove that Stark resonances for the Coulomb potential 
are defined on the whole complex plane based on our complex distortion.

We set $P=-\Delta +x_1+V(x)$ and assume Assumption~\ref{asm-potential}. 
We define the distorted operator $P_{\theta}$ from $P$ as in Section 2. 
The distortion is performed outside $\mathrm{supp}\,V_{\mathrm{sing}}$. 
\begin{lemma}\label{lem-appendix}
$V_{\mathrm{sing}}$ is $P_{\theta}$-bounded with relative bound $0$.
\end{lemma}
\begin{proof}
Set $P_{1, \theta}=P_{\theta}-V_{\mathrm{sing}}$. 
We take a cutoff function $\chi \in C_c^{\infty}(\RR^n)$ near $\mathrm{supp}\,V_{\mathrm{sing}}$. 
Assumption~\ref{asm-potential} implies that for $u \in  C_c^{\infty}(\RR^n)$ 
and for any small $\ve>0$ 
\[
\|V_{\mathrm{sing}}u\|=\|V_{\mathrm{sing}}\chi u\| 
\le \ve \|-\Delta \chi u\|+C_{\ve}\|\chi u\|.
\]
Since $P_{1,\theta}=-\Delta+x_1+V_1(x)$ and 
$x_1+V_1(x)$ is bounded near $\mathrm{supp}\,\chi$, 
\begin{align*}
\|V_{\mathrm{sing}}u\|
&\le \ve \|P_{1, \theta}\chi u\|
 +C_{\ve}\|u\|\\
&\le \ve \|P_{1, \theta}u\|+C_{\ve}\|u\|
 +\ve\|[-\Delta, \chi]u\|\\
&\le 2\ve \|P_{1, \theta}u\|
 +C_{\ve}\|u\| \\
&\le 2\ve \|P_{\theta}u\|
 +2\ve\|V_{\mathrm{sing}}u\|
 +C_{\ve}\|u\|, 
\end{align*}
where the third inequality follows from the standard elliptic estimate. 
Subtracting $2\ve\|V_{\mathrm{sing}}u\|$ for small $\ve>0$, the proof is finished.
\end{proof} 

\begin{proof}[Proof of Theorem~\ref{thm-0}]
We first prove \cite[Proposition 2.1]{K} in this case, that is, 
$P_{\theta}$ is an analytic family of type (A) and $P_{\theta}^*=P_{\bar{\theta}}$. 
As in \cite[subsection 2.1]{K}, it is enough to prove 
\begin{equation}\label{eq-appendix}
\|(P_{\theta}-P_{\theta'})u\|\le C |\theta-\theta'|\|P_{\theta}u\|+C_{\theta, \theta'}\|u\| \tag{A.1}
\end{equation}
for $u \in C_c^{\infty}(\RR^n)$. 
By the case where $V_{\mathrm{sing}}=0$, 
we have 
\begin{align*}
\|(P_{\theta}-P_{\theta'})u\|&=\|(P_{1,\theta}-P_{1,\theta'})u\|\\
&\le C |\theta-\theta'|\|P_{1,\theta}u\|+C_{\theta, \theta'}\|u\| \\
&\le C |\theta-\theta'|\|P_{\theta}u\|
 +C|\theta-\theta'|\|V_{\mathrm{sing}}u\|+C_{\theta, \theta'}\|u\|.
\end{align*}
This and Lemma~\ref{lem-appendix} imply the inequality~(\ref{eq-appendix}).

We next prove \cite[Proposition 2.2]{K} in this case, 
that is, $P_{\theta}$ with $\mathrm{Im}\theta<0$ has purely discrete spectrum in 
$\{z|\, \mathrm{Im}z>\mathrm{Im}\theta\}$.  
As in \cite[subsection 2.2]{K}, it is enough to prove that for 
$\mathrm{Im}z>\mathrm{Im}\theta$ and large $M>1$, we have 
$\|(\widetilde{P_{\theta}}-z)u\|\ge c \|u\|$ 
for $u\in \Cinf_c(\RR^n)$.  
Here 
\[
\widetilde{P_{\theta}}=P_{\theta}-iM\phi(x/M)\phi(D/M)^2\phi(x/M), 
\]
where $\phi \in C_c^{\infty}(\RR^n)$, $0\le \phi \le 1$, $\phi=1$ near $\{|x|\le 1/3\}$, 
$\mathrm{supp}\phi \subset \{|x|\le 1\}$ and $\int_{\RR^n} \phi(x)dx =1$. 
We fix small $\varepsilon_1>0$ and set $\chi_{j,M}=\tau_j(G(x)/M)$, where 
$\tau_0 \in C_b^{\infty}(\mathbb{R})$ is a cutoff near $(-\infty, \varepsilon_1]$,  
$\tau_1 \in C_b^{\infty}(\mathbb{R})$ is a cutoff near $[\frac{1}{2}\varepsilon_1, \frac{3}{2}\varepsilon_1]$,
$\tau_2 \in C_b^{\infty}(\mathbb{R})$ is a cutoff near $[2\varepsilon_1, \infty)$ and 
$G(x)=(1+K^{-2})^{\frac{1}{2}}\mathrm{dist}\left(\bullet,  \widetilde{C}(K, 0)\right)\ast \phi$, 
where $\phi$ is as above. 
We fix $z$ with $\mathrm{Im}z>\mathrm{Im}\theta$. 
We set 
$Q=\widetilde{P_{\theta}}-V_{\mathrm{sing}}-z+\chi_{2,M}w-iM\chi_{2,M}$ 
and define the symbol $q$ by $Q=q(x, D; M)$. 
Here $w$ is the same as in Section~4. 
Denote the seminorms in $S(\langle \xi \rangle^k)$ 
by $|a|_{k,\alpha}=
\sup_{x,\xi}|\partial_{x, \xi}^{\alpha}a|/\langle \xi \rangle^k$. 
We have 
\[
\left|\frac{\langle\xi\rangle^{2-k}}{q(x, \xi; R)}\right|\le CM^{-k/2}
\]
for $0\le k \le 2$. 
This is proved if we estimate them separately for 
$|x|<M/3$, $x\in C(K, 3\ve_1 M)^c$ and $x_1/M\lesssim 1$ for small $\ve_1>0$ 
and for $|\xi|/M^{1/2}\lesssim 1$ and $|\xi|/M^{1/2}\gg 1$ as in \cite[Proposition 2.2]{K}. 
Then we have 
\begin{align*}
\left|\frac{\partial_x q(x, \xi; M)}{q(x, \xi; M)}\right| 
\lesssim \left|\frac{\langle\xi\rangle^2}{q(x, \xi; M)}\right|\lesssim 1
\end{align*}
and 
\begin{align*}
&\left|\frac{\partial_{\xi}q(x, \xi; M)}
{q(x, \xi; M)}\right|\lesssim
\left|\frac{\langle\xi\rangle}{q(x, \xi; M)}\right|\lesssim M^{-1/2}.
\end{align*}
Thus $|q^{-1}|_{k-2, \alpha}=\mathcal{O}(M^{-k/2})$ for $0\le k \le 2$ and 
$|\partial_{\xi}q^{-1}|_{-2, \alpha}=\mathcal{O}(M^{-1/2})$ 
(see Remark~\ref{correction} below).
Since $\partial_{\xi}q\in S(\langle \xi \rangle)$ and $\partial_x q\in S(\langle \xi \rangle^2)$, 
the estimates $|\partial_x q^{-1}|_{-1, \alpha},\,\, |\partial_{\xi}q^{-1}|_{-2, \alpha}
=\mathcal{O}(M^{-1/2})$ are enough to estimate 
the seminorms in $S(1)$ of the symbols of $q^{-1}(x, D; M)q(x, D; M)-1$ and $q(x, D; M)q^{-1}(x, D; M)-1$. 
By the Neumann series argument and Beals's theorem, we thus conclude that 
$Q^{-1}=\mathcal{O}(M^{-k/2})$ in $\mathrm{Op}S(\langle \xi \rangle^{k-2})$ for large $M$ and 
$0 \le k \le 2$. 

Thus we have 
\begin{align*}
M\|\chi_{0, M}u\|&\le C\|Q\chi_{0, M}u\|
=C\|(\widetilde{P_{\theta}}-V_{\mathrm{sing}}-z)\chi_{0, M}u\| \\
&\le C\|(\widetilde{P_{\theta}}-z)\chi_{0, M}u\|
 +C\| V_{\mathrm{sing}}\chi_{0, M}u\|\\
&\le C\|(\widetilde{P_{\theta}}-z)\chi_{0, M}u\|
 +\ve C\|(P_{\theta}-z)\chi_{0, M}u\|
 +C_{\ve}\|\chi_{0, M}u\|,
\end{align*}
where the last inequality follows from Lemma~\ref{lem-appendix}. 
We take $\ve <\frac{1}{4C}$ and then take $M>2C_{\ve}$. 
Subtracting $C_{\ve}\|\chi_{0, M}u\|\le \frac{M}{2}\|\chi_{0, M}u\|$, we have 
\begin{align*}
\|\chi_{0, M}u\|&\le C\|(\widetilde{P_{\theta}}-z)\chi_{0, M}u\|+\frac{2\ve C}{M}\|(P_{\theta}-z)\chi_{0, M}u\|\\
&\le C\|(\widetilde{P_{\theta}}-z)\chi_{0, M}u\|+\frac{1}{2M}\|iM\phi(x/M)\phi(D/M)^2\phi(x/M)\chi_{0, M}u\|\\
&\le C\|(\widetilde{P_{\theta}}-z)\chi_{0, M}u\|+\frac{1}{2}\|\chi_{0, M}u\|. 
\end{align*}
Subtracting $\frac{1}{2}\|\chi_{0, M}u\|$, we have $\|\chi_{0, M}u\|\le C \|(\widetilde{P_{\theta}}-z)\chi_{0, M}u\|$. 

The remaining part of the proof of \cite[Proposition 2.2]{K} for $V_{\mathrm{sing}}\not \equiv 0$ 
is essentially the same as in \cite[subsection 2.1]{K} with minor modifications as follows. 
We set $\widetilde{\chi}_{0, M}=1-\chi_{0, M}$. 
Note that $V_{\mathrm{sing}}=0$ near 
$\mathrm{supp}\chi_{1, M}$, $\mathrm{supp}\chi_{2, M}$ and $\mathrm{supp}\widetilde\chi_{0, M}$.
We have 
\[
-\mathrm{Im}(\widetilde{\chi}_{0, M} u, (\widetilde{P_{\theta}}-z)\widetilde{\chi}_{0, M} u)\ge c\|\widetilde{\chi}_{0, M} u\|^2 
\]
for large $M>1$ by the complex distortion outside a cone. 
Thus we have 
\begin{align*}
 \|u\|&\le  \|\chi_{0, M} u\|+\|\widetilde{\chi}_{0, M} u\|\\
            & \le C\|(\widetilde{P_{\theta}}-z)\chi_{0, M} u\|+C\|(\widetilde{P_{\theta}}-z)\widetilde{\chi}_{0, M} u\| \\
			&\le C\|(\widetilde{P_{\theta}}-z)u\|+C \|[\widetilde{P_{\theta}}, \chi_{0, M}]u\|. 
\end{align*}		
We have 
\begin{align*}
\|[\widetilde{P_{\theta}}, \chi_{0, M}]u\|
\le CM^{-1}\|\chi_{1,M} u\|_{H^1}+\mathcal{O}(M^{-\infty})\|u\|. 
\end{align*}
Since $\mathrm{supp}\,\chi_{1,M}\cap\mathrm{supp}\,V_{\mathrm{sing}}=\emptyset$, we have 
\begin{align*}
\|\chi_{1,M} u\|_{H^1}
&\le C\|Q\chi_{1,M} u\|_{H^{-1}}\\
&=C\|(\widetilde{P_{\theta}}-z)\chi_{1,M}u\|_{H^{-1}} \\
&\le C\|(\widetilde{P_{\theta}}-z)u\|+
C\|[\widetilde{P_{\theta}},\chi_{1,M}] u\|_{H^{-1}}\\
&\le C\|(\widetilde{P_{\theta}}-z)u\|+CM^{-1}\|u\|. 
\end{align*}
Summing up these, we have 
\[
\|u\|\le C \|(\widetilde{P_{\theta}}-z)u\|+CM^{-2}\|u\|. 
\]
By subtracting $CM^{-2}\|u\|$ from both sides, 
we have $\|(\widetilde{P_{\theta}}-z)u\|\ge c \|u\|$ for large $M>1$.   

Once Proposition 2.1 and Proposition 2.2 in \cite{K} are proved, 
the proof of Theorem~\ref{thm-0} is the same as that of \cite[Theorem 1]{K}.
\end{proof}

\begin{remark}\label{correction}
The statement that $|q^{-1}|_{-2,\alpha}=\mathcal{O}(M^{-|\alpha|)/2})$ 
in the proof of \cite[Proposition 2.2]{K} was too strong. 
We correct it here. 
The necessary argument for the modification is straightforward and contained in the above proof. 
Namely, weaker estimate for $|q^{-1}|_{-2, \alpha}$, $|q^{-1}|_{-1, \alpha}$ 
and $|\partial_{\xi}q^{-1}|_{-2, \alpha}$ are proved similarly and are enough for the following arguments.  
\end{remark}

Resonances for $P$ coincide with discrete eigenvalues of $P_{\theta}$ 
in the region $\{z|\,\mathrm{Im}z>\mathrm{Im}\theta\}$ for $\mathrm{Im}\theta<0$ including multiplicities. 
This is proved by the same proof as in \cite[Section~2]{K}. 

\begin{remark}
The other results in \cite[Section~2]{K} are also true in the almost same form 
under Assumption~\ref{asm-potential} by the same proofs as in \cite{K}. 
For instance, we may replace $L^p_{\mathrm{comp}}$ by 
$L^p_{\mathrm{cone}}=\{f \in L^p | \mathrm{supp}f \subset C(K, \rho) \mspace{7mu} 
\text{for some} \mspace{7mu} K, \rho\}$ in our Theorem~\ref{thm-0}.
There is some modifications concerning the unique continuation argument. 
Namely, we assumed $\chi_j=1$ near $\mathrm{supp}\,V_{\mathrm{sing}}$ in our Theorem~\ref{thm-0} 
and should assume $U\supset \mathrm{supp}\,V_{\mathrm{sing}}$ in \cite[Proposition~2.3]{K}. 
If we moreover assume that 
there is a closed set $S \subset \mathbb{R}^n$ of Lebesgue measure zero such that 
$\mathbb{R}^n\setminus S$ is connected and 
$V_{\mathrm{sing}}$ is bounded on any compact subset of $\mathbb{R}^n\setminus S$, 
then these modifications are not necessary. 
For \cite[Theorem 3]{K}, the same proof shows the existence of 
a one-to-one correspondence between the shape resonances of $P$ 
and eigenvalues of a reference operator $P^{\mathrm{int}}$ 
with their distances bounded by $e^{-S/{\hbar}}$. 
Thus \cite[Theorem 3]{K} holds true if the Weyl law for $P^{\mathrm{int}}$ is true. 
\end{remark}

\section*{Acknowledgement}
The author is grateful to Shu Nakamura and Kenichi Ito for encouragement.
The author is supported by JSPS KAKENHI Grant Number JP21J10860.

Graduate School of Mathematical Sciences, 
the University of Tokyo, 3-8-1, Komaba, Meguro-ku, Tokyo 153-8914, Japan

E-mail address: kameoka@ms.u-tokyo.ac.jp

\end{document}